\documentclass[preprint,3p,times]{elsarticle}

\usepackage{amssymb,amsmath}
\usepackage{float,graphicx,lipsum}
\usepackage{siunitx}
\DeclareSIUnit{\molar}{M}
\usepackage{soul}
\usepackage{url}
\usepackage[skip=4pt]{caption} 
\usepackage{multirow}
\usepackage{booktabs}
\usepackage{pifont}
\usepackage{hyperref}
\usepackage{pdfpages}

\usepackage[draft]{changes}
\definechangesauthor[name={Simon Ternes}, color=blue]{ST}
\definechangesauthor[name={Giulio Barletta}, color=green]{GB}

\biboptions{sort&compress}
\usepackage{breakurl}

\usepackage[dvipsnames]{xcolor}

\usepackage{enumitem}
\newlist{todolist}{itemize}{2}
\setlist[todolist]{label=$\square$}
\usepackage{pifont}

\makeatletter
\let\els@orig@fnmark\fnmark
\def\fnmark[#1]{%
  \edef\@tempa{#1}\edef\@tempb{samecont}%
  \ifx\@tempa\@tempb
    \begingroup
      \renewcommand\thefootnote{\fnsymbol{footnote}}%
      \textsuperscript{,}\footnotemark[2]%
    \endgroup
  \else
    \els@orig@fnmark[#1]%
  \fi
}
\makeatother

\usepackage{lineno}

\journal{}

\begin{document}

\begin{frontmatter}

\title{
Quantifying Perovskite Solar Cell Degradation via Machine Learning from Spatially Resolved Multimodal Luminescence Time Series
}

\author{Giulio Barletta$^{1,\dagger}$}
\author{Simon Ternes$^{2,\dagger}$}
\author{Saif Ali$^{2}$}
\author{Zohair Abbas$^{2}$}
\author{Chiara Ostendi$^{2}$}
\author{Marialucia D'Addio$^{2}$}
\author{Erica Magliano$^{3}$}
\author{Pietro Asinari$^{1,4}$}
\author{Eliodoro Chiavazzo$^{1,}$\corref{cor1}}
\author{Aldo Di Carlo$^{2,3,}$\corref{cor2}}

\address{$^{1}$ Department of Energy, Politecnico di Torino, Corso Duca degli Abruzzi, 24, Turin, 10129, Italy}
\address{$^{2}$ CHOSE---Centre for Hybrid and Organic Solar Energy, University of Rome Tor Vergata, Via del Politecnico, 1, Rome, 00133, Italy}
\address{$^{3}$ ISM-CNR, Institute of Structure of Matter, Consiglio Nazionale delle Ricerche, Via Fosso del Cavaliere, 100, Rome, 00133, Italy}
\address{$^{4}$ INRIM, Istituto Nazionale di Ricerca Metrologica, Strada delle Cacce, 91, Turin, 10135, Italy}

\cortext[cor1]{\href{mailto:eliodoro.chiavazzo@polito.it}{eliodoro.chiavazzo@polito.it}}
\cortext[cor2]{\href{mailto:aldo.dicarlo@uniroma2.it}{aldo.dicarlo@uniroma2.it}}
\begingroup
  \renewcommand\thefootnote{\fnsymbol{footnote}}%
  \footnotetext[2]{Authors contributed equally}%
\endgroup

\begin{abstract}
Perovskite solar cells achieve remarkable power conversion efficiencies, yet operational stability remains a major barrier to large-scale deployment.
Reliable and rapid assessment of device state of health is therefore essential.
Conventional electrical diagnostics, such as illuminated current-voltage (J--V) sweeps, provide accurate performance metrics but are time-consuming and do not resolve spatially localized degradation, motivating non-invasive imaging-based alternatives.
A deep-learning framework is introduced to estimate PSC efficiency retention, $R_\mathrm{PCE}=\mathrm{PCE}_t/\mathrm{PCE}_0$, directly from multimodal luminescence imaging acquired during device aging.
Each sample combines electroluminescence (EL), open-circuit photoluminescence (PL\textsubscript{oc}), and short-circuit photoluminescence (PL\textsubscript{sc}) at an aged state with device-specific reference images at $t=0$, enabling learning of degradation-relevant spatial changes.
LumPerNet, a compact convolutional neural network, is benchmarked against a spatially homogenized control in which each luminescence channel is replaced by its spatial average while retaining the same learning framework and leakage-aware protocol.
The comparison indicates that global luminescence evolution contains most of the predictive signal, while spatial information provides a secondary contribution to robustness.
These results establish spatially resolved luminescence imaging as a practical route for accelerated stability testing and non-invasive degradation monitoring in perovskite photovoltaics.
\end{abstract}


\end{frontmatter}


\newpage
\section{Introduction}\label{sec:introduction}
Perovskite solar cells (PSCs) have emerged as one of the most promising emerging photovoltaic (PV) technologies~\cite{green_solar_2025,kojima_organometal_2009,green_emergence_2014,jung_perovskite_2015,kim_high-efficiency_2020,correa-baena_promises_2017}, achieving a remarkable leap in power conversion efficiency (PCE), from around 3.8\% in 2009 to over 27\% for single junction PSCs~\cite{xiong_homogenized_2025,noauthor_longi_2024} at the time of writing.
Furthermore, the flexible bandgap of perovskite absorber thin films has enabled their deployment in the tandem device architecture, yielding 34.85\% for perovskite--Si tandem cells~\cite{noauthor_3485_2025}, and 30.1\% for perovskite--perovskite tandem cells in 2025~\cite{lin_all-perovskite_2025}.
Despite these performance milestones, their medium- to long-term operational stability remains a critical challenge, with reports demonstrating stable operation on the order of $10^3$~h under standardized stress protocols, although full $T_{80}$ values are not always reached or consistently reported~\cite{khenkin_consensus_2020,wang_review_2019,mokabane_review_2025,yaghoobi_nia_co-crystal_2026}.
Unlike silicon PV, which have demonstrated moderate degradation over three decades defining the industry standard~\cite{fazal_progress_2023,cui_technoeconomic_2022,chen_technology_2023}, PSCs are highly sensitive to environmental stressors such as moisture, oxygen, UV light, and thermal cycling.
On the short to medium timescale (up to $10^2$~h), the performance of perovskite based PV was shown to fluctuate according to mostly reversible transient effects such as light soaking and path dependence~\cite{bian_unraveling_2025,lin_light_2023}.
However, on the medium to long timescale ($>10^2$~h), irreversible degradation of the perovskite crystal structure to lead halide (rich) phases is routinely observed~\cite{ahmad_instability_2017,kalasariya_how_2026,dubose_hole_2022,guo_unveiling_2025}.
This instability is mediated by intrinsic material vulnerabilities (including phase transitions, ion migration, and interfacial degradation) as well as extrinsic factors like poor encapsulation and crystallographic defects~\cite{zhu_long-term_2023,mokabane_review_2025,ahmedchowdhury_stability_2023,ahmad_instability_2017,wang_review_2019,kim_material_2016,silverman_durability_2025}.

While stability has historically been one of the main barriers for perovskite PV, recent years have witnessed significant progress toward addressing these limitations: for instance, early outdoor field trials have demonstrated stable operation despite the technology's relative youth~\cite{jiang_towards_2023,jost_perovskite_2020,meheretu_recent_2024}.
Furthermore, several studies document thousands of hours of continuous operation under realistic conditions and under standardized stress conditions similar to those applied to silicon~\cite{jacobsson_open-access_2022,nur-e-alam_current_2025,azmi_damp_2022,mavlonov_thermal_2024,lintangpradipto_single-crystal_2023}, such as elevated-temperature storage or damp heat (ISOS D2 and D3 elevated stress tests~\cite{khenkin_consensus_2020,reese_consensus_2011}).
However, translating these stability advances into rapid iteration and scalable quality control still requires monitoring tools that are faster than repeated full electrical characterization and that can localize where degradation initiates.
This motivates non-invasive luminescence imaging as a practical readout for accelerated testing, where data-driven models can translate spatially resolved patterns into device-level performance retention even when the target metric is spatially averaged, as explored in this work.

The operational stability of metal-halide perovskites is governed by coupled chemical and ionic processes that are strongly influenced by the environment and by electrostatic boundary conditions within the device stack~\cite{khenkin_consensus_2020,wang_review_2019}.
Operational stability drifts can be broadly grouped into reversible (metastable) and irreversible contributions, which may coexist and act on different time scales~\cite{khenkin_consensus_2020,vkhenkin_reconsidering_2018}.
A central irreversible failure route is the partial or complete conversion of the perovskite absorber into PbI$_2$ (and volatile organic species in MA/FA-based compositions), which can be triggered by moisture-assisted hydrolysis and hydrate formation, photo-oxidation in the presence of oxygen, and elevated temperature, each of which perturbs the ionic lattice and facilitates defect formation and ion motion~\cite{wang_review_2019,aristidou_role_2015}.
Starting at grain boundaries or the interface with other materials, the decomposition of the perovskite absorber into PbI$_2$ and volatile components propagates into the bulk, facilitated by ion mobility and the ingress of humidity and/or oxygen in the perovskite crystal structure~\cite{sun_role_2017,manekkathodi_observation_2020,tian_grain_2021,aristidou_role_2015}.
At the device level, both irreversible and reversible performance changes are closely linked to mobile ionic defects (notably halide-related vacancies and interstitials) that redistribute under illumination and bias, leading to internal field screening, interfacial ion accumulation, and time-dependent changes in recombination, hysteresis, and apparent performance~\cite{khenkin_consensus_2020,thiesbrummel_ion-induced_2024}.
In mixed-halide compositions, ionic redistribution can also manifest as light-induced halide segregation, with partial reversibility upon returning to dark conditions~\cite{kalasariya_how_2026}.
In addition to degradation, repeated or continuous bias and/or illumination (``light soaking'') can induce temporary performance improvements, often attributed to evolving interfacial electrostatics and contact properties mediated by ionic redistribution rather than permanent bulk passivation~\cite{kress_persistent_2022,moghadamzadeh_spontaneous_2020,shao_elimination_2016}.
Consistently, partial or even near-complete recovery of device performance after dark storage has been widely reported and is commonly referred to as self-healing, highlighting the metastable character of several degradation modes in PSCs~\cite{finkenauer_degradation_2022,prete_bias-dependent_2021,kalasariya_how_2026}.
At interfaces, irreversible degradation can be amplified by interfacial redox chemistry and compositional exchange, including electrode- or TCO-related elemental migration into the absorber or transport layers, which can alter band alignment and increase interfacial recombination and contact barriers~\cite{ahmad_instability_2017,wang_review_2019}.
Device-level stressors therefore map onto distinct degradation signatures:
humidity primarily accelerates chemical decomposition and interfacial reactions~\cite{azmi_damp_2022,song_efficient_2025};
UV and high-energy photons can promote interfacial contact degradation and, in oxide-based architectures, photocatalytic pathways~\cite{chen_ultraviolet_2022,lee_uv_2016,farooq_spectral_2018};
thermal stress and thermal cycling exacerbate diffusion, interconnect weaknesses, and contact non-uniformities~\cite{mavlonov_thermal_2024,li_direct_2017};
electrical bias can further drive ion accumulation at interfaces and promote irreversible reactions~\cite{qi_recent_2025,bae_electric-field-induced_2016,prete_bias-dependent_2021}.
To summarize, multiple cause-effect relationships have been reported between specific stressors, changes in material structure, and device-level performance variations.
Although recent studies have proposed data-driven approaches to predict performance dynamics under different stress conditions~\cite{kouroudis_artificial_2024,chakar_bridging_2025,dunlap-shohl_physiochemical_2024}, the large number of relevant variables, the diversity of device stacks and encapsulation strategies, and the scarcity of exhaustive degradation-transient datasets still make accurate prediction challenging~\cite{ali_outdoor_2023}.
This is expected given the high ion mobility and phase metastability of hybrid perovskites compared with classical semiconductors~\cite{thiesbrummel_ion-induced_2024,ahn_towards_2024}, as well as the relative youth of the technology compared with established PV platforms such as silicon~\cite{noauthor_best_nodate}.
For the first commercial roll-outs of perovskite PV~\cite{jowett_oxford_2024,kennedy_plenitude_2025,shaw_chinese_2024}, close monitoring of field performance is therefore particularly important.

For the aforementioned reasons, developing robust, non-invasive methods to monitor device degradation transients and predict the associated performance decline is integral to realizing the full potential of PSCs for utility-scale deployment.
Beyond providing a stability figure of merit, monitoring degradation through time enables practical decisions during both development and deployment.
For instance, it can (i) support high-throughput screening of materials, interfaces, and encapsulation strategies by rapidly identifying device architectures that develop localized defects, (ii) guide targeted follow-up diagnostics by pinpointing where degradation initiates (e.g., edges, contacts, shunted regions), and (iii) enable automated quality control and early-warning indicators in pilot-scale or field operation, where frequent electrical characterization is impractical.

The standard approach to monitoring PV performance relies on classical electrical characterization techniques such as current--voltage (J--V) curves, external quantum efficiency (EQE), and impedance spectroscopy.
While these methods provide an accurate characterization of device performance in laboratories working on research and development, they are relatively time-consuming and require measurement setups capable of load sweeping and/or standardized illumination.
For these reasons, classical characterization is poorly suited for rapid performance assessments on large-area PV.
Therefore, faster and less complex characterization tools are important for studying degradation in utility-scale PV deployments and for rapid screening in high-throughput fabrication lines.
Moreover, conventional electrical characterization does not provide spatial resolution, limiting its ability to identify local failure points such as degradation hotspots or shunt pathways.
In silicon PV, electroluminescence (EL) and photoluminescence (PL) imaging are now well-established for detecting cracks, shunts, hotspots, and potential-induced degradation, and have been successfully scaled to field-level inspections through drone-based and lock-in techniques~\cite{puranik_progress_2023,wang_drone-based_2024,doll_photoluminescence_2021}.
Recently, these image-based diagnostic techniques have gained traction as non-invasive tools for detecting and quantifying degradation also in perovskite PV~\cite{puranik_progress_2023,chen_imaging_2019,soufiani_electro-_2016,mariam_environmentally_2023,schall_accelerated_2023}.
Still, as compared to silicon PV, the use of EL and PL imaging for PSCs remains comparatively underexplored.
While PL was extensively used in laboratory studies to probe carrier lifetimes, trap densities, and interfacial recombination~\cite{campanari_reevaluation_2022,geng_can_2023,kim_interfacial_2021,kruckemeier_understanding_2021,dileep_k_rapid_2021}, systematic efforts to link the temporal evolution of luminescence patterns during aging with device degradation are still scarce~\cite{hacene_intensity_2024,basumatary_probing_2022,chen_carrier_2021}.
In contrast, EL imaging of PSCs is less explored due to additional challenges arising from the required electrical injection of charges such as ion migration, contact instability, and inter-device variability, which complicate the interpretation of spatial patterns~\cite{schall_accelerated_2023,wu_anomalous_2024,torre_cachafeiro_ion_2025}.
%

Existing machine learning (ML)- driven approaches in PSC diagnostics can be broadly grouped into three classes.
First, a large body of work uses EL/PL imaging (often at a single time point) for static characterization tasks, such as defect segmentation~\cite{gao_definition_2023,louis_operando_2025,taherimakhsousi_quantifying_2020,ternes_correlative_2022} or inference of electrical parameters and absolute efficiency from spatial patterns~\cite{laufer_process_2023,laufer_deep_2025,kumar_machine_2025,zhang_machine_2025}.
Second, complementary ML efforts target lifetime and stability metrics using tabular descriptors of composition, processing, and operating conditions, typically leveraging literature or database-derived labels rather than operando-imaging time series~\cite{odabacsi_machine_2020,hu_machine-learning_2022,zhao_machine_2025,mammeri_stability_2025}.
Third, recent studies have begun to exploit time series of optical measurements (including high-throughput or operando luminescence) with data-driven forecasting or self-supervised learning, highlighting the potential of temporal optical signals for monitoring degradation dynamics~\cite{srivastava_machine_2023,ji_self-supervised_2023}.

Despite these advances, only a few recent studies have begun to connect perovskite luminescence imaging with data-driven prediction of electrical performance or degradation-related behavior.
Glaws et al.~\cite{glaws_explainable_2025} used transfer learning and explainable AI to predict current–voltage metrics from different combinations of EL and PL images, identifying which imaging conditions are most informative for $V_\mathrm{oc}$, $I_\mathrm{sc}$, FF, and PCE.
Schall et al.~\cite{schall_using_2026} extended this direction to operando PL image sequences during accelerated stress testing, using a CNN–LSTM model with attention and Grad-CAM attribution to predict $V_\mathrm{oc}$ transients and analyze electrical metastability.
These studies demonstrate the relevance of luminescence-based ML diagnostics for perovskite devices, but differ from the present work in the predicted target and input formulation.
Here, we focus on direct prediction of the device-level efficiency-retention metric $R_\mathrm{PCE}$ during aging by combining repeated EL, PL\textsubscript{oc}, and PL\textsubscript{sc} imaging with device-specific reference-current pairing and a leakage-aware device-level validation protocol.

The proposed methodology departs from traditional degradation modeling approaches that are mainly based on empirical extrapolation~\cite{zhao_accelerated_2022,pandey_progress_2026,ye_degradation_2023} or isolated electrical measurements~\cite{tirawat_measuring_2024,hauff_impedance_2022}.
Instead, it leverages multimodal time series of luminescence images, combining EL and PL data acquired at both reference and aged states, to train ML models that regress degradation directly from visual data.
In this work, we adopt the efficiency retention $R_\mathrm{PCE}$, defined as the ratio between the instantaneous and initial PCE
\begin{equation} 
    {R_\mathrm{PCE}}(t) = \frac{\mathrm{PCE_{avg}}(t)}{\mathrm{PCE_{avg}}(t=0)},
    \label{eq:r_pce} 
\end{equation}
as a degradation metric for PSCs.
Here, the $\text{PCE}(t)$ is the PCE recorded at time $t$ computed as the average of forward and reverse J--V scans.
This quantity is often used in literature as a measure of performance degradation~\cite{vkhenkin_reconsidering_2018,he_amorphous_2026,song_efficient_2025}.
Although it has been criticized as an absolute measure for degradation studies due to the possibility of hiding prior degradation in the normalization constant~\cite{khenkin_consensus_2020,vkhenkin_reconsidering_2018,tiihonen_critical_2018}, it serves here as the target variable to let all investigated targets start at the same value at $t=0$ for our predictive framework.
By explicitly combining temporal dynamics with complementary luminescence modalities, the framework is designed to provide a non-invasive route for quantifying degradation without explicit electrical measurements at the device terminals.

In detail, we demonstrate, on a representative dataset spanning distinct device architectures, that convolutional neural networks (CNNs) trained on multimodal time series of luminescence images can learn degradation-relevant spatial patterns and predict the efficiency retention of PSCs with sufficient accuracy for reliable monitoring and screening.
While EL and PL probe distinct recombination pathways and can respond differently to spatially localized defects, it is not yet clear whether a reduced set of modalities would suffice.
We therefore explicitly quantify the benefit of combining EL, PL\textsubscript{oc}, and PL\textsubscript{sc} through an exhaustive ablation study over all single- and paired-modality inputs.
This work focuses on establishing a general and extensible framework for image-based $R_{\text{PCE}}$ prediction, rather than on optimizing predictive accuracy for a specific dataset.

Such an image-driven approach can complement established electrical diagnostics and outperform space-averaged monitoring by exploiting automated, spatially resolved, and scalable monitoring of device aging.
It could support accelerated stability testing, production-line quality control, and eventually field-level durability screening.
More broadly, integrating $R_\mathrm{PCE}$ prediction with luminescence imaging offers a scalable data-driven approach to monitor and screen degradation in PSCs.
While not intended to replace mechanistic models, such tools can guide targeted experimental characterization to elucidate dominant failure mechanisms, a prerequisite for achieving long-term stability in next-generation PV technologies.

This work is structured as follows.
Section~\ref{sec:methods} describes the employed methodology, starting with the description of the experimental automated aging and imaging setup, continuing with the end-to-end dataset construction, from region of interest (ROI)-based pre-processing to $R_{\text{PCE}}$ label generation, and ending with the introduction of the cross-validated (CV) training protocol and device-level held-out testing.
Section~\ref{sec:results} reports quantitative and qualitative results for multimodal image-based $R_{\text{PCE}}$ prediction, including the comparison to a spatially homogenized control baseline, and the modality-ablation study.
Finally, Section~\ref{sec:conclusions} summarizes the main findings, discusses limitations, and outlines directions toward full-lifecycle and deployment-oriented monitoring.
%
%
\section{Methods}\label{sec:methods}
\begin{table}[h!]
\centering
\caption{Notation used for multimodal luminescence imaging and efficiency-retention prediction.}
\label{tab:notation}
\begin{tabular}{p{0.18\linewidth} p{0.76\linewidth}}
\toprule
Symbol & Meaning \\
\midrule
$t$ & Aging time (elapsed time since the start of the aging experiment). \\
$0$ & Reference condition, i.e., data acquired for a specific device at $t=0$ before aging. \\
$\mathrm{EL}(t)$ & Electroluminescence image acquired at aging time $t$. \\
$\mathrm{PL}_{\mathrm{oc}}(t)$ & Photoluminescence image acquired at aging time $t$ under open-circuit conditions (oc). \\
$\mathrm{PL}_{\mathrm{sc}}(t)$ & Photoluminescence image acquired at aging time $t$ under short-circuit conditions (sc). \\
$r_\mathrm{EL}(t)$ & Pixel-wise ratio channel for EL, defined as $\mathrm{EL}(t) / \mathrm{EL}(0)$. \\
$r_\mathrm{PL_{oc}}(t),\, r_\mathrm{PL_{sc}}(t)$ & Pixel-wise ratio channels $\mathrm{PL}_{\mathrm{oc}}(t)/\mathrm{PL}_{\mathrm{oc}}(0)$ and $\mathrm{PL}_{\mathrm{sc}}(t)/\mathrm{PL}_{\mathrm{sc}}(0)$. \\
$\mathbf{x}(t)$ & Stacked multimodal input tensor at time $t$ (see Eq.~\ref{eq:tensor}). \\
$\mathrm{PCE}(t)$ & Power conversion efficiency measured from indoor J--V at aging time $t$. \\
$R_\mathrm{PCE}$ & Efficiency retention, $R_\mathrm{PCE}=\mathrm{PCE}(t)/\mathrm{PCE}(0)$. \\
$\hat{y}$ & Model prediction of $R_\mathrm{PCE}$ for a given sample. \\
\bottomrule
\end{tabular}
\end{table}

\subsection{Device fabrication}\label{ssec:fabrication}
This work investigates batches of perovskite solar cells with two distinct device architectures.
The first architecture A, CsFAMA PSCs fabricated by a two-step hybrid (thermal evaporation / solution) process, constitutes the majority of devices in the dataset and was tested in three separate batches, which share the same layer sequence and composition but underwent aging characterization at different times.
The second architecture B, solution-processed FACs PSCs, is used as a second example to enhance the generalization capability of the employed methodology.
In both cases the inverted p-i-n structure is used.

In the dataset bookkeeping, four experimental batch IDs are used to denote independent fabrication and aging/imaging campaigns.
Batch 0, Batch 1, and Batch 3 correspond to Architecture A devices measured in three separate campaigns, whereas Batch 2 corresponds to Architecture B devices.
These experimental batch IDs are metadata labels and should not be confused with neural-network mini-batches used during stochastic optimization.
In the following, we report the fabrication protocols for Architectures A and B.

\subsubsection{Substrate preparation and hole transport layer (HTL) deposition}
Flourine-doped tin oxide (FTO) coated glass for Architecture A and indium tin oxide (ITO) coated glass for Architecture B was cut into $\SI{2.5}{} \times \SI{2.5}{\centi\meter\squared}$ substrates, which were then patterned by P1 isolation using an ultraviolet picosecond laser, in order to define four independent pixels per substrate.
Substrates were sequentially cleaned with detergent solution, deionized water, acetone, and 2-propanol (\SI{10}{\minute} sonication each), followed by \SI{15}{\minute} UV/ozone treatment.
The substrates were transferred to a nitrogen filled-glovebox for HTL deposition.
\SI{80}{\micro\liter} of a \SI{0.5}{\milli\gram\per\milli\liter} solution of (4-(3,6-Dimethoxy-9H-carbazol-9-yl)butyl)phosphonic acid (MeO-4PACz) in ethanol for Architecture A or (2-(3,6-Dimethoxy-9H-carbazol-9-yl)ethyl)phosphonic acid (MeO-2PACz) in ethanol for Architecture B were deposited on the glass substrates and, following a \SI{10}{\second} waiting time, spin coated at 5000 rpm for \SI{30}{\second} and then annealed at \SI{100}{\degreeCelsius} for \SI{10}{\minute}.

\subsubsection{Interlayer, perovskite deposition and passivation for Architecture A}
After the HTL deposition, a \SI{80}{\micro\liter} of a \SI{0.5}{\milli\gram\per\milli\liter} solution of 2-decyl[1]benzothieno[3,2-b][1]benzothiophene (C10-BTBT) in dimethylformamide (DMF) was spin coated at 5000 rpm for \SI{20}{\second} and annealed at \SI{100}{\degreeCelsius} for \SI{5}{\minute}.

For the two-step hybrid deposition, the substrates were loaded in a multi-source thermal evaporation system (KE-500K by Kenosistec) integrated within a nitrogen-filled glovebox.
The pressure in the chamber was brought to \mbox{$\sim10^{-6}$~mbar}, a film of PbI\textsubscript{2} and CsBr (with a ratio of 10:1) was thermally co-evaporated onto the substrates.
Afterwards, formamidinium iodide (FAI; 0.48 M), methylammonium bromide (MABr; 0.09 M), and methylammonium chloride (MACl; 0.09 M) were dissolved in ethanol and dynamically spin-coated at 4500 rpm for \SI{30}{\second}, in order to form a $\mathrm{Cs_{0.12}FA_{0.79}MA_{0.09}Pb(I_{0.85}Br_{0.15})_3}$ perovskite. 
The devices were then annealed in ambient air, with a controlled humidity in the range $30\%-40\%$, at \SI{125}{\degreeCelsius} for \SI{25}{\minute}.
The devices were brought to a dry-air glovebox where a post-treatment was performed by spin coating phenethylammonium chloride (PEACl; \SI{1.5}{\milli\gram\per\milli\liter} in ethanol) at 5000 rpm for \SI{30}{\second} and annealing at \SI{100}{\degreeCelsius} for \SI{10}{\minute}.

\subsubsection{Perovskite deposition and passivation for Architecture B}
The perovskite precursor solution (\SI{1.4}{\molar} $\mathrm{Cs_{0.17}FA_{0.83}Pb(I_{0.75}Br_{0.25})_3}$) was prepared in $N,N$-dimethylformamide (DMF) and dimethyl sulfoxide (DMSO) (DMF:DMSO = 4:1 v/v) and stirred overnight at room temperature.
The perovskite layer was deposited by spin coating (\SI{100}{\micro\liter}, 4000 rpm, \SI{35}{\second}); chlorobenzene (CB, \SI{150}{\micro\liter}) was used as antisolvent during the last \SI{15}{\second}), followed by annealing at \SI{100}{\degreeCelsius} for \SI{20}{\minute}.
Post-treatment was performed by spin coating phenethylammonium chloride (PEACl; \SI{2}{\milli\gram\per\milli\liter} in isopropyl alcohol (IPA)) at 6000 rpm for \SI{30}{\second} and annealing at \SI{100}{\degreeCelsius} for \SI{5}{\minute}.

\subsubsection{Electron transport layer (ETL), buffer layer, and top contact deposition}

The electron transport layer phenyl-C61-butyric-acid-methyl ester (PCBM; \SI{27}{\milli\gram\per\milli\liter} in CB:dichlorobenzene (DCB) = 0.75:0.25 v/v, \SI{80}{\micro\liter}) was deposited by dynamic spin coating at 2000 rpm for \SI{30}{\second} and annealed at \SI{100}{\degreeCelsius} for \SI{5}{\minute}.
An aluminium-doped zinc oxide (AZO) layer was deposited after PCBM by spin coating as reported in Ref.~\cite{magliano_solution-processed_2025}.
A \SI{170}{\nano\meter} ITO top electrode was then sputtered using a Kenosistec KS-400 sputtering system. For Architecture B only, a \SI{100}{\nano\meter} copper layer was deposited by thermal evaporation through a shadow mask which protected the active area of each pixel. 
Finally, for each pixel, a portion of the positive FTO bottom contact was uncovered by removing the perovskite layer with a 7:1 DMF:xylene solution, and silver paste was applied to both the bottom FTO (or ITO for Architecture B) and top ITO contacts.

\subsection{Justification of input features}\label{ssec:physics}

Electroluminescence (EL), open-circuit photoluminescence (PL\textsubscript{oc}) and short-circuit photoluminescence (PL\textsubscript{sc}) were selected as input modalities to the proposed ML framework because they probe complementary aspects of carrier generation, recombination, injection, and extraction in operating PSCs.
Although all three signals originate from radiative recombination, they are acquired under different (opto-)electrical operation states and therefore emphasize different device-physics contributions to the measured luminescence pattern.
In the literature, the device physics are typically related to the external luminescence yield, $0<\eta<1$, by
\begin{equation}
V_{\mathrm{OC}} \approx V_{\mathrm{OC, rad}} - \Delta V_{\mathrm{OC, non-rad}}  = \frac{k_\mathrm{B} T}{e} \ln \left( \frac{J_\mathrm{SC}}{e \Phi_{\mathrm{ em, 0}}} \right) - \frac{k_\mathrm{B} T}{e} \ln \left(\frac{1}{\eta} \right) 
\label{eq:voc_rec}
\end{equation}
where $V_{\mathrm{OC}}$ is the open-circuit voltage of the device, $V_{\mathrm{OC, rad}}$ the maximum achievable open-circuit voltage only limited by radiative recombination (radiative limit), $ \Delta V_{\mathrm{OC, non-rad}}$ is the open circuit loss introduced by non-radiative recombination at interfaces and the bulk of the absorber and $\Phi_{\mathrm{ em, 0}}$ is the total emitted photon flux of the solar cell that is related to its (spectral) absorption coefficient due to reciprocity of emission and absorption~\cite{tress_perovskite_2017}.
The quantum yield $\eta$ is sometimes set equal to the external EL quantum efficiency ($\mathrm{EQE}_{EL}$)\cite{tress_perovskite_2017}, or analogously, equal to the external PL quantum yield under optical excitation (external $\mathrm{PLQY}$)~\cite{wolff_nonradiative_2019}.
Thus, both $\mathrm{EQE}_{EL}$ and external $\mathrm{PLQY}$ probe the ratio of radiative to non-radiative recombination and both EL and PL intensities are expected to correlate with the device performance.
However, in practice, this is not always observable in experimental measurements, especially when acquiring large-area images including in- and out-coupling related artifacts~\cite{thiesbrummel_understanding_2023}.
In large-area EL images, variation in charge carrier conductivity over interfaces as well as lateral and series resistances can have a higher impact than in PL images~\cite{wu_anomalous_2024}, where the absorber is probed through direct optical excitation, potentially obtaining more response from poorly contacted areas of the device~\cite{regalado-perez_impact_2021}.
Therefore, both EL and PL imaging modes can yield complementary information.
In addition, the necessity of using different conditions of electrical bias during PL measurements was recently stressed~\cite{campanari_reevaluation_2022}.
The reason is the strong non-ideality in perovskite solar cells and bias-dependent transient effects related to ion drifts during device operation~\cite{du_p-doping_2019,fassl_effect_2019}. 

Choosing PL\textsubscript{sc}, PL\textsubscript{oc}, and EL images as input features is fundamentally motivated by their ease of measurement over large areas.
Importantly, they can be measured rapidly and nearly non-invasively with high spatial resolution over areas of \SI{}{\meter\squared}, without requiring electrical load sweeping or resistance measurements, as would be needed for maximum-power-point characterization.
This makes them suitable candidates for fault detection in large utility-scale PV deployment and for rapid quality screening in high-throughput production, where fast large-scale assessments of PV module state of health are important~\cite{puranik_progress_2023,chen_imaging_2019,soufiani_electro-_2016,mariam_environmentally_2023,schall_accelerated_2023}.
The spatial resolution further enables the detection of patterns such as degradation hotspots and morphological defects, which may not yet be reflected in J--V scans but can enhance the predictive capability of the developed model.

We note that this deliberate choice to prioritize ease of measurement over complete optoelectrical characterization can cause information loss.
First, the measured camera intensity integrates multiple optical and device effects, including absorber thickness, excitation in-coupling, emission out-coupling, optical interference, charge transport, and contacting.
Therefore, a high measured luminescence intensity does not necessarily correspond to a higher value of $\eta$.
Second, one camera pixel corresponds to an active area of approximately \SI{0.1}{\milli\meter\squared}, so microscopic effects such as grain-dependent degradation cannot be resolved.
Nevertheless, for the large-area industrial use cases, there is value in predicting device performance and its evolution from these simple multimodal measurements.

\subsection{Device aging and monitoring}\label{ssec:monitoring}
\begin{figure}[h]
    \centering
    \includegraphics[width=\linewidth]{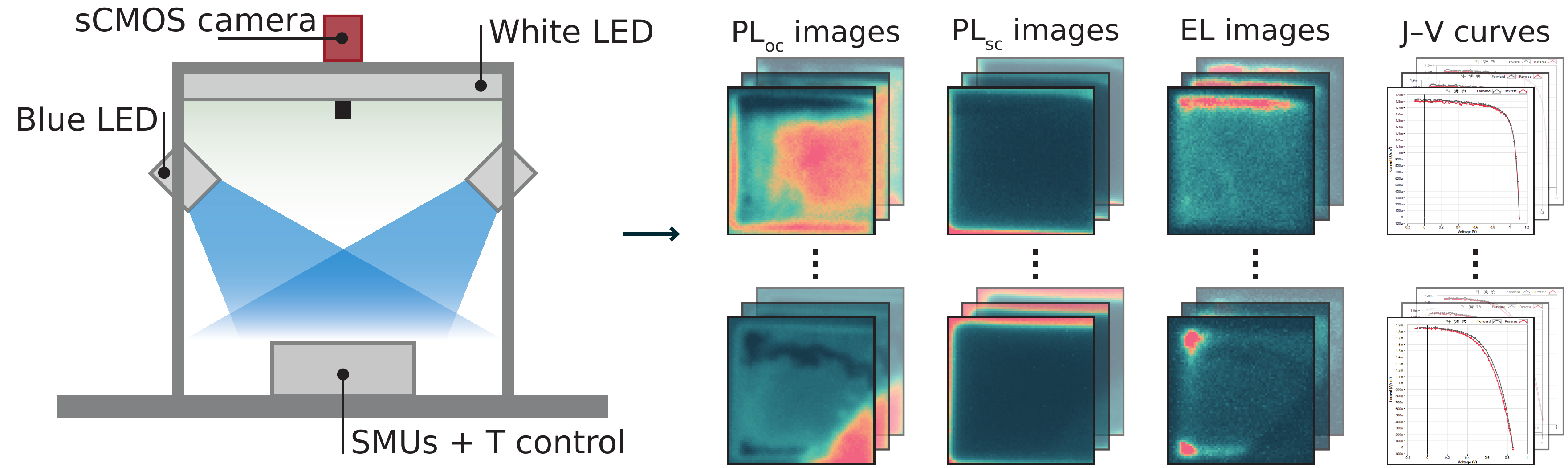}
    \caption{
    Experimental monitoring workflow and acquired data modalities.
    A custom chamber integrates temperature control and source-measure units (SMUs) with two illumination channels (white LED for J--V acquisition, blue LED for PL excitation) and an sCMOS camera for spatially resolved imaging.
    At user-selected aging times, the automated monitoring cycle yields paired PL images at open circuit ($\mathrm{PL}_{\mathrm{oc}}$) and short circuit ($\mathrm{PL}_{\mathrm{sc}}$), EL images under forward bias (\SI{1.5}{V}), and corresponding J--V curves, forming multimodal time series for each device. 
    }
    \label{fig:workflow}
\end{figure}

Stability tests were conducted in a custom-built setup designed to enable controlled degradation of PSCs with automatic integrated temperature control, electrical characterization and luminescence imaging.
The experimental monitoring workflow is summarized in Figure~\ref{fig:workflow}, including a schematic of the custom chamber and instrumentation (left) and the resulting multimodal time series acquired at successive aging times (right), namely $\mathrm{PL}_{\mathrm{oc}}$, $\mathrm{PL}_{\mathrm{sc}}$, EL images, and indoor J--V curves.
All hardware components were coordinated through a Raspberry Pi 4 (Ubuntu OS), which managed device communication and task synchronization via Python-based control scripts.
The single-board computer issued digital control signals through GPIO pins to drive illumination sources, triggered imaging acquisition over USB 3.0, and interfaced with the electrical characterization and aging platform via its application programming interface (API).
Electrical biasing and current-voltage characterization were performed using the Arkeo system (Cicci Research srl), which enabled both J--V measurements and device biasing during luminescence acquisition.
Indoor J--V curves were recorded under broadband white illumination provided by a DLP Ring Light (Smart Vision Lights).
Note that the white-LED J--V measurements are used to compute relative efficiency retention under consistent indoor illumination rather than to report absolute AM1.5G-certified efficiencies.
Unless otherwise stated, all electrical performance metrics reported in this work (V\textsubscript{oc}, J\textsubscript{sc}, fill factor (FF), and PCE) refer to these indoor J--V measurements acquired under controlled laboratory illumination.
For EL imaging, a forward bias of \SI{1.5}{V} was applied to inject carriers and induce radiative recombination.
For PL imaging, the system sequentially imposed open-circuit and short-circuit conditions while the devices were optically excited by two LSR300 linear light bars (470 nm LEDs, Smart Vision Lights), mounted on aluminum supports to ensure uniform excitation across the devices, mechanical stability, and effective heat dissipation during continuous operation.
Normalized emission spectra of the white and blue LED sources, as well as a photograph of the experimental setup, are provided in the Supplementary Information (Figures~S1 and S2).

Luminescence images were captured using a Quantalux CS2100M-USB sCMOS camera (Thorlabs) equipped with a FELH0600 long-pass filter ($\lambda_{\mathrm{cut-on}}=\SI{600}{\nano\meter}$, Thorlabs) to suppress the excitation light and isolate the emission signal.
Camera exposure time and gain were kept constant throughout the test series for each imaging mode to enable quantitative comparison of luminescence intensity.
Each image contained up to 32 cells arranged in a fixed layout (see Section~\ref{ssec:processing}).

\begin{figure}
    \centering
    \includegraphics[width=0.8\linewidth]{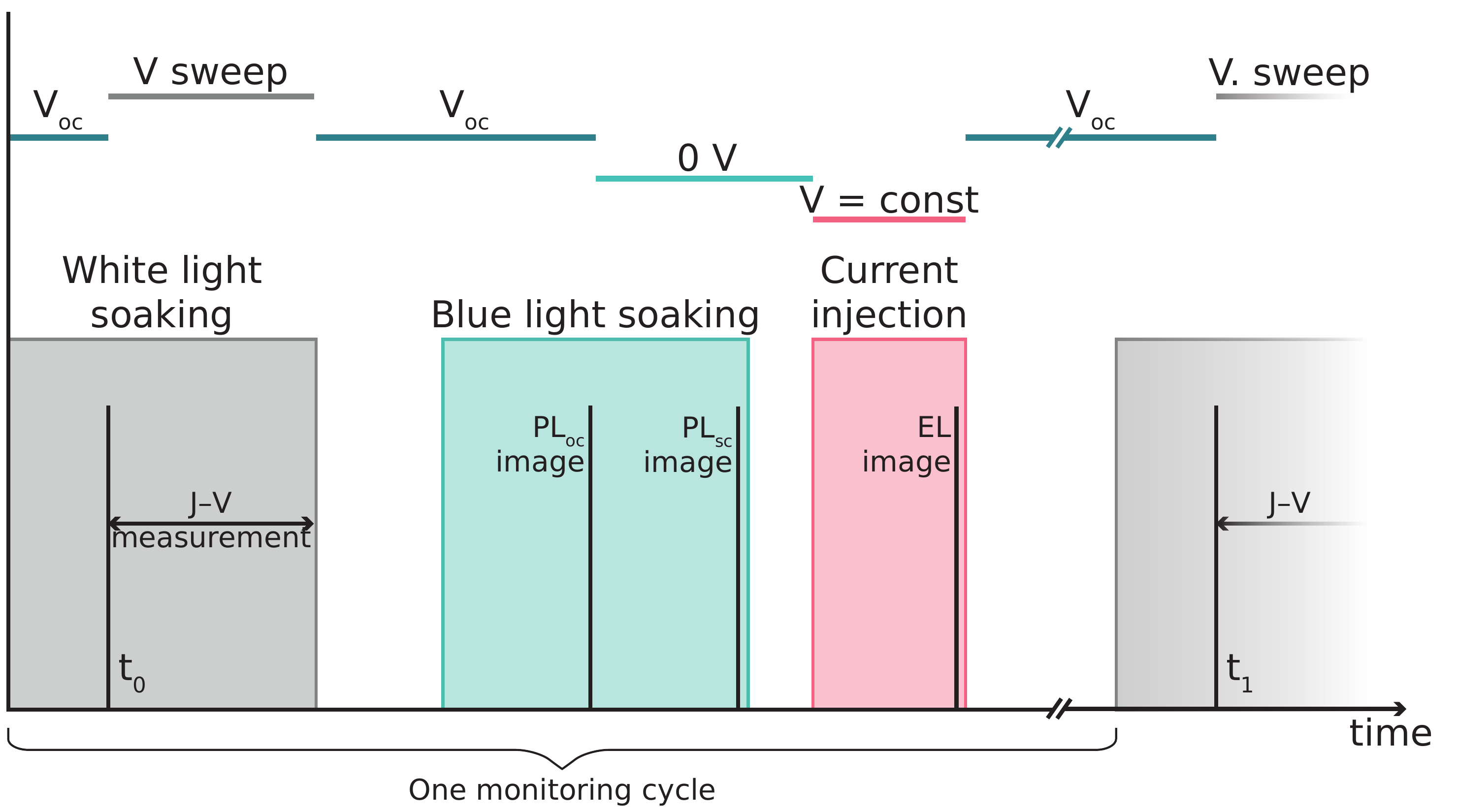}
    \caption{Timing diagram of one automated monitoring cycle executed between two consecutive user-selected aging times, $t_0$ and $t_1$.
    The cycle comprises (i) white-light soaking (\SI{20}{\second}) followed by a J--V sweep under white illumination (\SI{30}{\second}); (ii) dark relaxation at open circuit (\SI{60}{\second}); (iii) PL imaging under continuous blue illumination at open circuit ($\mathrm{PL}_{\mathrm{oc}}$, \SI{10}{\second}) and then immediately at short circuit ($\mathrm{PL}_{\mathrm{sc}}$, \SI{10}{\second}; \SI{20}{\second} total blue illumination); (iv) dark relaxation at short circuit (\SI{15}{\second}); and (v) EL imaging during forward-bias current injection (\SI{10}{\second}).
    J--V sweeps were performed at every cycle, whereas luminescence images were stored only at a subset of cycles following a progressively sparser schedule at later aging times; the illumination and voltage program of the cycle was otherwise unchanged across all iterations.}
    \label{fig:cycle}
\end{figure}

Devices under analysis were kept at a constant temperature, controlled through the Arkeo platform and maintained at \SI{30}{\degreeCelsius} in ambient atmosphere for the proof-of-concept aging series reported herein. The devices were not encapsulated to enable ingress of humidity to increase performance variation during the measurement time (in future measurements, humidity will be exactly controlled and monitored). The temperature was not elevated above \SI{30}{\degreeCelsius} to prevent photoluminescence quenching.
At user-defined time intervals (identical across all batches), the control software executed a fully automated measurement sequence (Figure~\ref{fig:cycle}) comprising (i) a J--V sweep under continuous white-LED illumination; (ii) PL imaging under continuous blue LED excitation at open-circuit ($V=V_{\mathrm{oc}}$) and short-circuit ($V=\SI{0}{\volt}$) conditions; and (iii) EL imaging under forward bias (\SI{1.5}{V}).
Within each cycle, all dwell times were kept constant across datasets: \SI{20}{\second} white-light soaking for recombination, \SI{30}{\second} white-light soaking for the J--V sweep, \SI{60}{\second} relaxation at open circuit in the dark to minimize bias-history artifacts, \SI{10}{\second} blue-light soaking for $\mathrm{PL}_{\mathrm{oc}}$ followed immediately by \SI{10}{\second} for $\mathrm{PL}_{\mathrm{sc}}$ (i.e., \SI{20}{\second} continuous blue illumination), \SI{15}{\second} dark relaxation at short circuit, and \SI{10}{\second} voltage bias for EL acquisition.
Between cycles, devices were held at open circuit in the dark to allow for device relaxation and self-healing.
J--V sweeps were performed at every monitoring cycle, images were stored according to a progressively sparser schedule to balance temporal resolution and acquisition overhead: images were recorded at every iteration for $\#_\mathrm{iter} \le 100$, every second iteration for $100 < \#_\mathrm{iter} \le 200$, and every fifth iteration for $\#_\mathrm{iter} > 200$.
Importantly, the illumination and bias steps associated with PL/EL were still executed at each cycle even when images were not recorded, thereby preserving identical boundary conditions and minimizing systematic differences between cycles with and without stored images.
This cycling routine allowed systematic tracking of device performance and luminescence evolution over aging time.
Figure~\ref{fig:cycle} schematically summarizes one full automated measurement cycle, highlighting the temporal sequence of electrical characterization and luminescence imaging steps.

\subsection{Data processing \& label generation}\label{ssec:processing}
Raw EL and PL image sequences acquired during the accelerated aging experiments were processed through a dedicated Python pipeline developed in-house.
The routine performs dark-field correction, hot-pixel removal, and spatial cropping of each device region of interest (ROI) prior to model training.
All operations were implemented in Python 3.12 using NumPy~\cite{harris_array_2020}, Pandas~\cite{mckinney_data_2010}, SciPy~\cite{virtanen_scipy_2020}, Matplotlib~\cite{hunter_matplotlib_2007}, and tifffile~\cite{christoph_gohlke_cgohlketifffile_2025}.
Each dataset folder contained time-indexed images from the three luminescence modalities (EL, PL\textsubscript{oc}, PL\textsubscript{sc}) acquired under identical spatial framing.
A single reference image was used to interactively define all ROIs corresponding to the individual cells tested within the same frame.
ROI coordinates were saved as JSON and CSV manifests, enabling reproducible batch processing without further user input.
Before cropping, each raw image was corrected for fixed-pattern artifacts through dark-field subtraction, following
\begin{equation}
    I_{\mathrm{corr}}=I_{\mathrm{raw}}-I_{\mathrm{dark}}.
    \label{eq:field_correction}
\end{equation}
A static hot-pixel mask was first derived from the dark frame itself to flag persistent defective pixels, while a second adaptive despiking step based on median-absolute-deviation statistics was applied frame-by-frame to suppress transient outliers.
Pixels exceeding ten standard deviations above the local median were winsorized rather than replaced outright, preserving image texture while mitigating radiometric spikes~\cite{russ_image_2016}.
After correction, each ROI was cropped into fixed-size ($72 \times 72$ px) patches centered on the predefined coordinates.
The cropped sub-images were stored in 16-bit TIFF format within a hierarchical folder structure organized by modality and channel, and a manifest file containing all processing metadata (modality, ROI ID, coordinates, and file paths) was automatically generated.

Raw luminescence and electrical data were processed through a two-stage Python workflow designed to ensure spatial and temporal consistency across all modalities.
In the first stage, current--voltage (J--V) logs exported from the Arkeo platform were parsed to extract, for each cell and time step, the forward and reverse power conversion efficiencies $\left( \mathrm{PCE_{fw}, PCE_{rv}} \right)$ together with the individual factors of the PCE, i.e., V\textsubscript{oc}, J\textsubscript{sc}, and FF.
These records were associated with the corresponding EL, PL\textsubscript{oc}, and PL\textsubscript{sc} images acquired during the same automated measurement cycles, yielding a unified manifest that maps each image triplet to its electrical response.
In the second stage, this manifest was used to assemble per-cell datasets for model training.
For each device, the luminescence images recorded at the initial cycle were designated as the reference state ($t=0$), and all subsequent images were paired with this baseline to form time-ordered sequences.
For every time step $t$, a multimodal tensor was constructed as
\begin{equation}
    \mathrm{\mathbf{x}}(t) = \left[ \mathrm{EL}(t), \mathrm{PL}_{\mathrm{oc}}(t), \mathrm{PL}_{\mathrm{sc}}(t), \mathrm{EL}(0), \mathrm{PL}_\mathrm{oc}(0), \mathrm{PL}_\mathrm{sc}(0), r_\mathrm{EL}(t), r_\mathrm{PL_{oc}}(t), r_\mathrm{PL_{sc}}(t) \right]
    \label{eq:tensor}
\end{equation}
where each $r(t)=I(t)/(I(0)+\varepsilon)$ represents a normalized intensity ratio with a small stabilizing constant $\varepsilon=10^{-6}$.
The multimodal tensor $x(t)$ is constructed by channel-wise concatenation of all selected image channels.
Therefore, data fusion is performed at the input level: EL, PL\textsubscript{oc}, PL\textsubscript{sc}, their corresponding $t=0$ reference images, and the ratio channels are jointly provided to the first convolutional layer rather than processed by separate modality-specific branches.

In this formulation, one supervised sample corresponds to one physical device observed at one stored aging time $t$ (the word ``sample'' is exclusively used in the sense of ``data sample'' herein, not ``laboratory sample'').
Thus, each device contributes multiple device-time samples, all sharing the same device-specific reference images acquired at $t=0$.
The reference state is fixed for a given device and is not redefined at later monitoring cycles.
The model therefore learns the degradation state from the joint representation of the current images, the initial reference images, and the corresponding pixel-wise ratio channels.
Unless otherwise stated, models receive the full multimodal tensor comprising EL, PL\textsubscript{oc}, and PL\textsubscript{sc} at both $t$ and $t=0$, together with the corresponding ratio channels.
For the modality-ablation experiments (Section~\ref{ssec:modalities}), we construct single- or double-modality tensors by retaining only the channels associated with the relevant modalities (e.g., $\mathrm{\mathbf{x}_{EL}}(t) = \left[ \mathrm{EL}(t), \mathrm{EL}(0), r_{\mathrm{EL}}(t) \right]$) and discarding the remaining channels.

Because each J--V measurement yields forward- and reverse-scan efficiencies, a single representative value was obtained by averaging the two,
\begin{equation}
    \mathrm{PCE_{avg}}(t) = \frac{1}{2} \left[ \mathrm{PCE_{fw}}(t) + \mathrm{PCE_{rv}}(t)\right],
\end{equation}
which provides a more robust and direction-independent estimate of device performance.
The degradation label was defined as the $R_\mathrm{PCE}$, calculated according to Equation~\ref{eq:r_pce}, representing the fraction of the original performance retained over time.
The analysis was restricted to samples with $R_\mathrm{PCE}$ values in the range 0.8--1.2, corresponding to operational regimes above end-of-life and excluding spikes in the early transient efficiency gains~\cite{khenkin_consensus_2020,zhang_big_2022,watts_light_2020,zhao_revealing_2015,tress_interpretation_2018}, while also limiting the regression target to the regime where sample density is sufficient for reliable training.
Each sample therefore combines temporally aligned multimodal imaging features with the corresponding quantitative degradation metric, forming a consistent basis for supervised learning of PSC aging.
All pre-processing steps preserve the spatial structure of the luminescence patterns and do not involve any learned or adaptive transformations.

\subsection{Machine learning framework}\label{ssec:framework}

\begin{figure}[t]
    \centering
    \includegraphics[width=\linewidth]{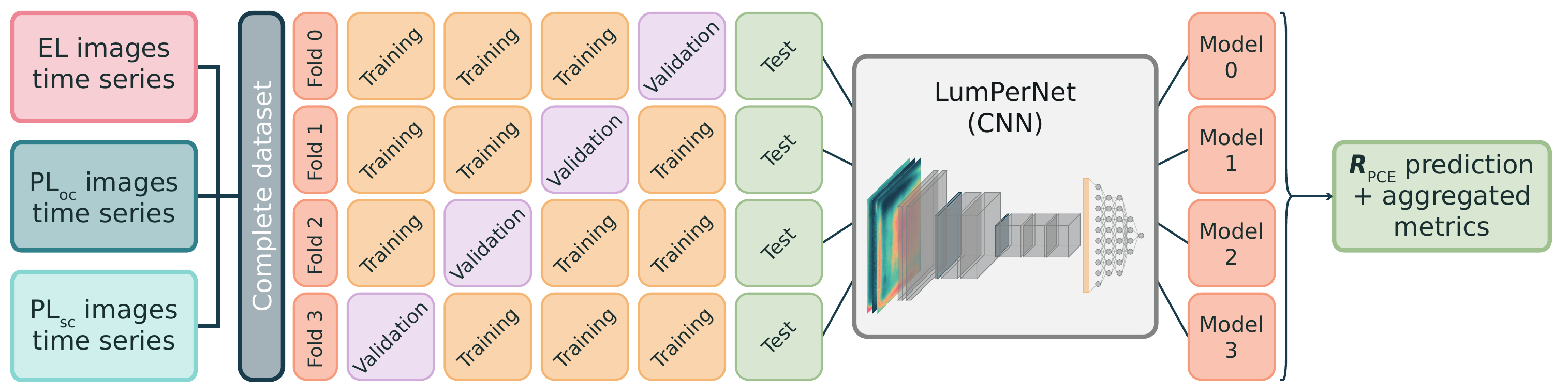}
    \caption{
    ML framework used in this work.
    Multimodal luminescence image time series (EL, $\mathrm{PL}_{\mathrm{oc}}$, $\mathrm{PL}_{\mathrm{sc}}$) constitute the model inputs.
    Two regressors are benchmarked:
    the full spatial LumPerNet, a CNN-based model operating on reference–current image pairs, and a spatially homogenized LumPerNet control in which each input channel is replaced by its spatial average before model ingestion.
    In LumPerNet, modality fusion is performed by early channel-wise concatenation of the selected EL, PL\textsubscript{oc}, and PL\textsubscript{sc} current, reference, and ratio channels, which are then processed jointly by the CNN backbone.
    Model selection is performed via four-fold cross-validation within the training--validation subset; for each fold, the best checkpoint is selected on the corresponding validation split and evaluated on the held-out test split.
    Final predictive performance is reported as the mean across folds, with the fold-to-fold standard deviation reported as an estimate of uncertainty.
    }
    \label{fig:ml_framework}
\end{figure}

The objective of the proposed ML framework is to predict the $R_\mathrm{PCE}$ of PSCs directly from multimodal luminescence imaging data acquired during device aging.
The problem is formulated as a supervised regression task, where each input sample consists of paired reference and current EL, PL\textsubscript{oc}, and PL\textsubscript{sc} images, and the target is the corresponding $R_\mathrm{PCE}$ value derived from electrical measurements.
The full workflow, including input modalities, baseline comparison, and the cross-validation protocol used for model selection and evaluation, is summarized in Figure~\ref{fig:ml_framework}.

To exploit both the spatial structure of luminescence patterns and their temporal evolution relative to the initial device state, we adopt a CNN-based architecture~\cite{lecun_deep_2015,li_survey_2022}, hereafter referred to as LumPerNet.
The image-based input tensor is initially processed to extract spatial features that encode degradation-related changes in EL and PL intensity distributions.
The latent representations produced by the CNN backbone are passed to a final Multilayer Perceptron (MLP) regressor~\cite{murtagh_multilayer_1991}, which outputs a scalar estimate of the device $R_\mathrm{PCE}$.
Importantly, the architecture operates on reference--current image pairs, enabling the model to learn relative changes over aging time rather than absolute intensity levels.
Batch IDs were used only for stratifying the device-level split and for the batch-conditioning ablation experiment reported in the Supplementary Information; the default LumPerNet model does not receive batch ID as an input feature.

To assess whether the proposed spatially resolved approach provides a tangible benefit beyond global luminescence evolution, we implemented a spatially homogenized control baseline.
In this baseline, each input channel of the multimodal tensor is replaced by its spatial average before being passed to the same LumPerNet learning framework.
This preserves the temporal, modality-specific, reference, current, and ratio information available to the full model, while removing spatial heterogeneity by construction.
The comparison therefore tests whether retaining spatially resolved luminescence patterns provides additional predictive value beyond global intensity evolution.

Model development and benchmarking followed a two-stage protocol designed to prevent information leakage and to enable fair comparisons across architectures.
First, the dataset was split at the device level into a training--validation subset ($\sim 80\%$ of total samples) and an independent held-out test subset ($\sim20\%$ of total samples), with stratification by device batch ID to preserve batch proportions across partitions.
The held-out test subset was not accessed during model development and was used only for final evaluation.

Within the training--validation subset, four-fold CV was performed, resulting in four independently trained models for each model architecture; importantly, the same train/validation/test splits were used consistently for both the full spatial LumPerNet and the spatially homogenized control, enabling a direct and controlled comparison in which the main controlled factor is the availability of spatial information.
After filtering to $R_\mathrm{PCE}\in[0.8,1.2]$, the dataset comprises $N_\mathrm{cells}=76$ physical devices and $N_\mathrm{samples}=13{,}138$ supervised device-time samples.
Here, a device-time sample denotes one stored monitoring time $t$ for one device, paired with that device's fixed $t=0$ reference state and labeled by the corresponding $R_\mathrm{PCE}(t)$.
The samples are distributed across $N_\mathrm{batches}=4$ experimental fabrication/measurement batch IDs, corresponding to the campaigns defined in Section 2.1, with 23, 23, 13, and 17 physical devices per batch, respectively.
All device-time samples originating from the same physical device were assigned to the same partition.
Therefore, no current image, reference image, or time point from a test device appears during training or validation.

For all folds, per-channel normalization statistics (mean and standard deviation) were computed on the training split only and then applied to validation and test samples.
During training, light geometric data augmentation was applied on-the-fly (small random translations and rotations, identical across all channels of a sample), and target-distribution imbalance within the selected $R_\mathrm{PCE}$ window was mitigated using weighted random sampling~\cite{efraimidis_weighted_2006}.
Reported metrics correspond to the mean and standard deviation of the performance obtained on the held-out test set across the four CV runs.
In addition, we also report committee (ensemble) performance obtained by averaging the predictions of the 4 fold-trained models on the pooled test set; these single-number metrics can differ from the mean of per-fold metrics.
The same splitting strategy and training and evaluation protocols are reused for all ablations to isolate the impact of input modality and spatial homogenization.
The training of both the full spatial LumPerNet and the spatially homogenized control, as well as the creation and handling of the dataset, were implemented in Python 3.12 using NumPy~\cite{harris_array_2020} and PyTorch~\cite{paszke_pytorch_2019}.
The exact split definitions and trained per-fold checkpoints are deposited together with the processed dataset on Zenodo, while the corresponding preprocessing, training, and evaluation scripts are provided in the public GitHub repository.

The comparison between the full spatial LumPerNet and the spatially homogenized control is intended as a proof-of-concept ablation to quantify the additional value of spatially resolved multimodal luminescence information beyond global intensity trends.
A comprehensive exploration of alternative architectures and hyperparameter optimization is left to future work.

%
%
\section{Results}\label{sec:results}

We test whether preserving spatially resolved EL/PL information improves prediction of the device-level, spatially averaged efficiency retention $R_\mathrm{PCE}$ beyond what is achievable from spatially homogenized multimodal luminescence inputs.
We evaluate efficiency retention prediction on time series of luminescence samples paired with synchronized J--V characterization, where each datapoint comprises steady-state multimodal images (EL, PL\textsubscript{oc}, PL\textsubscript{sc}) at an aged state $t$ together with device-specific reference images acquired at $t=0$, and is labeled with $R_\mathrm{PCE}(t)$ as defined in Section~\ref{ssec:processing}.
The analysis is restricted to $R_\mathrm{PCE} \in [0.8, 1.2]$, focusing on operational efficiency-retention regimes above end-of-life and excluding early transient efficiency gains.
Model development follows the leakage-aware protocol in Section~\ref{ssec:framework}, comprising a device-level split into training--validation and fixed held-out test subsets and four-fold CV within the training--validation subset; metrics are reported as mean $\pm$ standard deviation on the held-out test set across folds.

\subsection{$R_\mathrm{PCE}$ prediction from multimodal luminescence}\label{ssec:prediction}
We first assess whether a compact CNN-based model (LumPerNet) can regress the device $R_\mathrm{PCE}$ directly from multimodal luminescence images paired with device-specific reference states.
All results in this section refer to the operational window $R_\mathrm{PCE} \in [0.8, 1.2]$ and to the evaluation protocol described in Section~\ref{ssec:framework} (held-out test set, four-fold CV on the remaining devices, and aggregation across four independently trained models).
The choice of this operational window is motivated by the density of recorded data points in this region ($>95\%$) during the three-day degradation experiments.
Across the held-out test set, LumPerNet achieved a mean absolute error of $\mathrm{MAE}=0.048\pm0.005$, a root-mean-square error of $\mathrm{RMSE}=0.061\pm0.006$, and a coefficient of determination $R^2=0.558\pm0.078$ (mean $\pm$ standard deviation across the four CV-trained models).
These results demonstrate that time series of EL/PL image pairs contain sufficient predictive signal to estimate quantitative $R_\mathrm{PCE}$ within the operational regime, even under a conservative device-level split designed to prevent leakage from temporally adjacent samples.

\begin{figure}[t]
    \centering
    \includegraphics[width=0.5\linewidth]{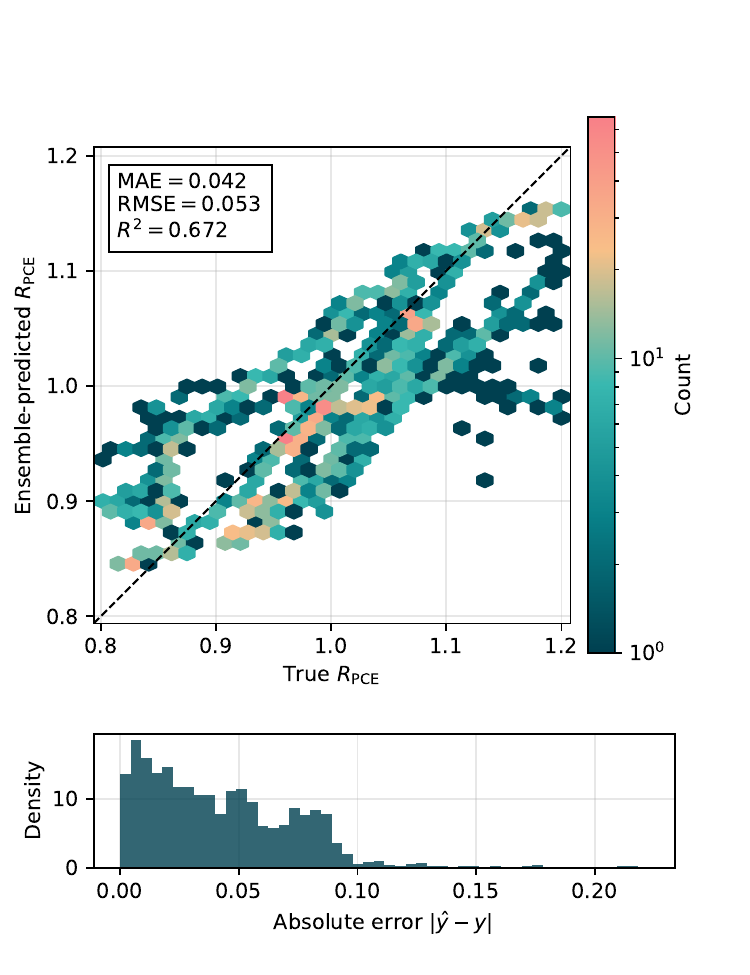}
    \caption{LumPerNet ensemble-mean parity plot and absolute error distribution for $R_\mathrm{PCE}$ prediction on the held-out test set.
    This ensemble-level performance is achieved from the full spatial LumPerNet; comparison with the spatially homogenized control is reported in Section~\ref{ssec:comparison}.
    Top: hexbin parity plot comparing measured $R_\mathrm{PCE}$ to the ensemble-mean prediction $\hat{y}$, computed by averaging the outputs of the four CV--trained LumPerNet models (one prediction per test sample).
    Color denotes the number of samples per bin, and the dashed line indicates the ideal $\hat{y}=y$ relationship; inset reports the corresponding ensemble performance metrics.
    Metrics shown in the inset are computed on the ensemble-mean predictor (mean of the 4 fold-trained models) evaluated on the held-out test set, whereas Table~\ref{tab:LumPerNet_nostack_perfold} reports mean $\pm$ std across fold-wise test evaluations.
    The visible banding around the diagonal reflects the longitudinal nature of the test set, with multiple measurements per device that are not statistically independent; device-specific residual structure can therefore emerge as correlated scatter.
    Bottom: distribution of absolute residuals $|\hat{y}-y|$ over the held-out test samples, summarizing the typical magnitude and spread of prediction errors within the operational range $R_\mathrm{PCE} \in [0.8, 1.2]$.
    }
    \label{fig:parity_hexbin}
\end{figure}

Performance was broadly consistent across folds.
Table~\ref{tab:LumPerNet_nostack_perfold} reports the per-fold metrics for both the validation and held-out test sets, together with the corresponding mean $\pm$ standard deviation across the four cross-validation runs.
The validation metrics quantify the sensitivity of model selection to the choice of held-out validation devices.
The lower validation $R^2$ values observed in some folds indicate split-dependent calibration difficulty during model selection.
However, these values should be distinguished from the final held-out test performance: the test set is identical across the four fold-trained models and contains devices disjoint from those used during training and validation.
On this fixed held-out test set, the fold-resolved $R^2$ values are substantially more stable, ranging from $0.490$ to $0.684$.
Figure~\ref{fig:parity_hexbin} visualizes predictions on the held-out test set using the ensemble-mean output obtained by averaging the four CV-trained models (one prediction per sample), together with the corresponding absolute error distribution.
For completeness, fold-resolved parity plots are reported in the Supplementary Information.

We additionally evaluated variants in which the models were conditioned on categorical batch identity through an embedding branch.
Here, ``batch'' denotes the fabrication and testing batch of each device.
Among the four batches considered, three correspond to devices with the same architecture measured in different campaigns, whereas one batch corresponds to a different architecture.
Explicit batch conditioning did not provide a robust improvement over the batch-free model and increased fold-to-fold variability in the per-fold metrics.
This suggests that the spatial luminescence representation already captures part of the batch-structured variability available in the images.
Because batch identity is dataset-specific and may not be available, or meaningful, for previously unseen fabrication or measurement domains, we retain the batch-free LumPerNet as the default formulation and report the batch-aware variants only as an ablation in the Supplementary Information.

\begin{table}[t]
\centering
  \caption{Per-fold performance of LumPerNet on the validation and held-out test sets ($R_\mathrm{PCE}$ prediction). Reported values are computed independently for each fold; the last row summarizes the mean $\pm$ standard deviation across folds.
  Each fold denotes one cross-validation training run performed on the device-level training--validation subset; the held-out test set contains disjoint devices and is identical across folds.}
  \label{tab:LumPerNet_nostack_perfold}
    \begin{tabular}{c ccc ccc}
        \toprule
        & \multicolumn{3}{c}{Validation} & \multicolumn{3}{c}{Test (held-out)} \\
        \cmidrule(lr){2-4} \cmidrule(lr){5-7}
        Fold & MAE $\downarrow$ & RMSE $\downarrow$ & $R^2$ $\uparrow$
             & MAE $\downarrow$ & RMSE $\downarrow$ & $R^2$ $\uparrow$ \\
        \midrule
        0 & 0.044 & 0.060 & 0.513 & 0.049 & 0.061 & 0.563 \\
        1 & 0.083 & 0.097 & 0.130 & 0.051 & 0.066 & 0.490 \\
        2 & 0.085 & 0.108 & 0.000 & 0.039 & 0.052 & 0.684 \\
        3 & 0.031 & 0.040 & 0.814 & 0.051 & 0.065 & 0.496 \\
        \midrule
        $\mu \pm \sigma$
          & 0.061 $\pm$ 0.024 & 0.076 $\pm$ 0.028 & 0.364 $\pm$ 0.321
          & 0.048 $\pm$ 0.005 & 0.061 $\pm$ 0.006 & 0.558 $\pm$ 0.078 \\
        \bottomrule
    \end{tabular}
\end{table}

\subsection{Baseline comparison: spatial features improve $R_\mathrm{PCE}$ prediction}\label{ssec:comparison}

To assess whether the predictive signal can be explained solely by global, space-averaged luminescence decay, we benchmark LumPerNet against an intensity-only baseline regressor.
The baseline model operates on space-averaged EL, PL\textsubscript{oc}, and PL\textsubscript{sc} intensities for each modality and time point, and therefore discards spatial information by construction.
LumPerNet, in contrast, learns spatial features from the full multimodal image tensor via convolutional layers.

\begin{figure}
    \centering
    \includegraphics[width=\linewidth]{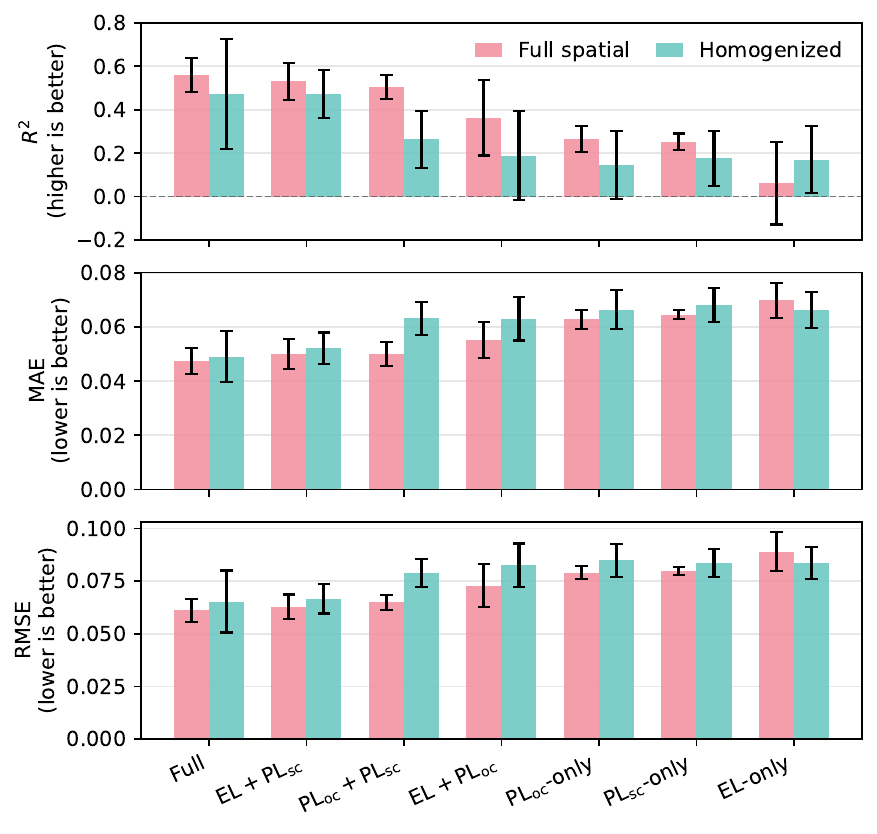}
    \caption{
    Cross-validated held-out test performance for full spatial LumPerNet and the spatially homogenized control under modality ablations.
    Bars report mean $\pm$ standard deviation across the four CV-trained models evaluated on the same fixed held-out test set.
    For each modality configuration, the full spatial model receives the corresponding current, reference, and ratio image channels, whereas the spatially homogenized control receives the same channels after replacing each image by its spatial average.
    The comparison shows that removing modalities generally degrades both models, but the performance loss is more pronounced for the homogenized control, indicating that spatially resolved information increases the robustness of the representation when reduced modality sets are used.
    Exact metrics are reported in Table~S2.}
    \label{fig:comparison_summary}
\end{figure}

Both models were trained and evaluated using identical device-level splits and the same two-stage protocol described in Section~\ref{ssec:framework}, enabling a direct and controlled comparison in which the main difference is the availability of spatial structure.
On the held-out test set, the spatially homogenized control achieved $\mathrm{MAE} = 0.049\pm 0.009$, $\mathrm{RMSE} = 0.065 \pm 0.015$, and $R^2 = 0.473 \pm 0.253$ (mean $\pm$ standard deviation across the four CV-trained models).
In comparison, as summarized in Figure~\ref{fig:comparison_summary} and Table~S1, the full spatial LumPerNet slightly improves test performance with lower fold-to-fold variability on the held-out test set under identical splits, particularly in $R^2$.

To further examine the role of spatial information, we quantified the spatial heterogeneity of each held-out test sample using the mean p90 -- p10 contrast of the EL, PL\textsubscript{oc}, and PL\textsubscript{sc} ratio channels.
This analysis, reported in the Supplementary Information, shows that spatial heterogeneity evolves during aging and that the full spatial LumPerNet tends to reduce fold-to-fold prediction dispersion relative to the spatially homogenized control for medium- and high-heterogeneity samples.
This supports the interpretation that spatial information contributes primarily to prediction robustness, while global luminescence evolution remains the dominant source of predictive signal.
These results indicate that global luminescence evolution contains a large fraction of the predictive signal, while preserving spatial structure provides a secondary contribution to robustness.

\subsection{Ablation on imaging modalities: EL, PL\textsubscript{oc}, and PL\textsubscript{sc} are complementary}\label{ssec:modalities}

Having compared the full spatial model against the spatially homogenized control, we subsequently quantify in an ablation study whether all three luminescence modalities are needed, or if a reduced input set suffices.
To this end, we repeated the full training and evaluation pipeline described in Section~\ref{ssec:framework} (device-level held-out test set, four-fold CV on the remaining devices, identical hyperparameters and model capacity) for both the full spatial LumPerNet and the spatially homogenized control while ablating the input tensor to retain
(i) a single modality at a time (EL-only, PL\textsubscript{oc}-only, or PL\textsubscript{sc}-only, still including the corresponding aged and reference images, as well as their ratio), and 
(ii) all coupled two-modality combinations (EL+PL\textsubscript{oc}, EL+PL\textsubscript{sc}, PL\textsubscript{oc}+PL\textsubscript{sc}).
The resulting model performance is summarized in Figure~\ref{fig:comparison_summary} and Table~S2.

Across all ablations, predictive accuracy degrades relative to the full multimodal LumPerNet model.
On the held-out test set, the full model reaches $\mathrm{MAE} = 0.048 \pm 0.005$ and $R^2 = 0.558 \pm 0.078$.
The single-modality variants show a clear hierarchy: EL-only performs weakest, with $\mathrm{MAE} = 0.070 \pm 0.007$ and $R^2 = 0.061 \pm 0.190$; in contrast, PL\textsubscript{oc}-only and PL\textsubscript{sc}-only retain more predictive information, reaching $\mathrm{MAE} = 0.063 \pm 0.003$ and $R^2 = 0.265 \pm 0.059$, and $\mathrm{MAE} = 0.065 \pm 0.002$ and $R^2 = 0.253 \pm 0.037$, respectively.
These results indicate that, in the present dataset, PL-based spatial patterns provide stronger standalone degradation descriptors than EL alone, while no single modality is sufficient to recover the full multimodal performance.

Coupling modalities partially recovers performance, but the gain depends strongly on \emph{which} physical contrasts are combined.
Among two-modality inputs, EL+PL\textsubscript{sc} performs best and is closest to the full model, achieving $\mathrm{MAE} = 0.050 \pm 0.006$ and $R^2 = 0.531 \pm 0.085$.
Interestingly, PL\textsubscript{oc}+PL\textsubscript{sc} is comparably competitive, with $\mathrm{MAE} = 0.050 \pm 0.005$ and $R^2 = 0.504 \pm 0.055$, showing that the two PL acquisition conditions already encode much of the information needed for robust $R_\mathrm{PCE}$ prediction.
By contrast, EL+PL\textsubscript{oc} performs worse ($\mathrm{MAE} = 0.055 \pm 0.007$, $R^2 = 0.364 \pm 0.175$) and exhibits the highest variability across folds.
This non-trivial behavior indicates that ``adding a second modality'' does not automatically increase robustness, and that some modality combinations can introduce competing cues that are not consistently transferable across device splits.

We further repeated the same modality-ablation protocol for the spatially homogenized LumPerNet control, in which each retained channel was replaced by its spatial average before model ingestion.
The paired comparison in Figure~\ref{fig:comparison_summary} shows that modality removal generally has a stronger adverse effect on the homogenized control than on the full spatial model, although the EL-only case remains an exception.
Thus, once spatial heterogeneity is removed, the model becomes more dependent on the simultaneous availability of all EL, PL\textsubscript{oc}, and PL\textsubscript{sc} global intensity signals.
Conversely, the full spatial LumPerNet is more tolerant to reduced modality sets, suggesting that spatially resolved patterns provide partially redundant degradation information that can compensate, to some extent, for missing global modality signals.
This supports the use of the full spatial multimodal input as the default configuration: although global luminescence evolution is highly informative, preserving spatial structure improves representation robustness under modality reduction.

A consistent interpretation emerges when considering the device physics of luminescence generation under different operating conditions.
First, both EL and PL\textsubscript{oc} maps locally probe radiative recombination under conditions of high quasi-Fermi-level splitting, although the excited carriers are generated by different mechanisms.
In EL, carriers are electrically injected, so the measured spatial pattern can be affected by contact quality, injection non-uniformity, ion migration, and local series resistance, in addition to recombination activity~\cite{soufiani_electro-_2016,wu_anomalous_2024}.
This may explain why EL alone provides limited standalone predictive performance in the present dataset.
By contrast, PL\textsubscript{oc} is optically excited and probes radiative recombination close to open-circuit conditions.
Despite the different carrier-generation mechanisms, both EL and PL\textsubscript{oc} are expected to correlate with the local implied V\textsubscript{oc}, because they reflect radiative recombination under high quasi-Fermi-level splitting~\cite{fischer_versatile_2025}.
This similarity also explains why the EL+PL\textsubscript{oc} ablation under-performs the other bimodal combinations: although it combines two acquisition modes, it lacks the complementary extraction-regime information provided by PL\textsubscript{sc}.
In contrast, PL\textsubscript{sc} operates in the carrier-extraction regime, where strong photoluminescence quenching indicates efficient extraction of charge carriers competing with radiative and non-radiative recombination channels~\cite{campanari_reevaluation_2022,wagner_revealing_2022}.
It therefore provides information complementary to both PL\textsubscript{oc} and EL~\cite{hameiri_photoluminescence_2015,wagner_revealing_2022}.
This interpretation is consistent with the strong performance of the EL+PL\textsubscript{sc} and PL\textsubscript{oc}+PL\textsubscript{sc} ablations, and suggests that PL\textsubscript{sc} supplies a particularly useful contrast for distinguishing degradation-relevant spatial changes.

Overall, these ablations support two practical conclusions.
First, the full multimodal representation provides the most robust overall generalization: the full model reduces test MAE by about 24\% compared to any single-modality alternative and yields a substantially higher $R^2$.
Second, reduced-input configurations can be competitive only when they preserve complementary physical information: a carefully chosen bimodal configuration (EL+PL\textsubscript{sc}, or alternatively PL\textsubscript{oc}+PL\textsubscript{sc}) can retain most of the performance of the full model, whereas EL+PL\textsubscript{oc} is not reliable under the present data splits.
The stronger degradation of the homogenized control under modality removal further indicates that spatial information stabilizes the representation when the global modality set is incomplete.
Taken together, these results justify the adoption of a multimodal input representation for robust $R_\mathrm{PCE}$ estimation within the operational window, and motivate reporting bimodal variants only when their cross-fold stability is explicitly verified.

\subsection{Qualitative analysis: time-evolving degradation examples}\label{ssec:qualitative}

\begin{figure}[t]
    \centering
    \includegraphics[width=\linewidth]{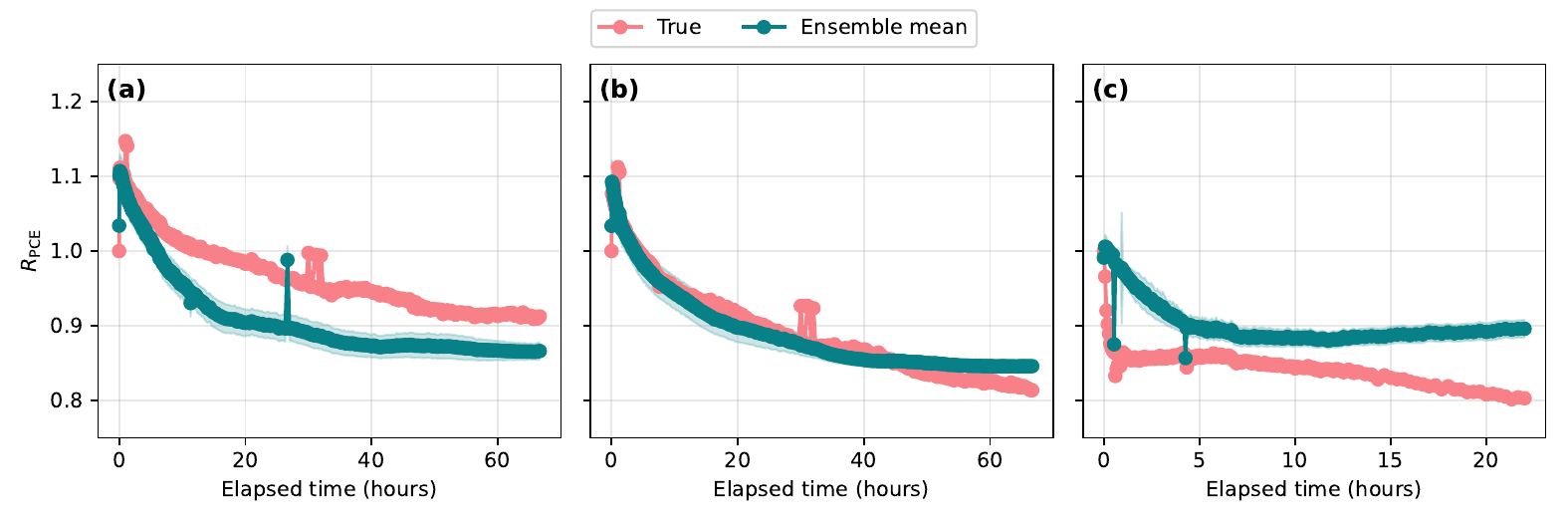}
    \caption{Representative time-evolving $R_\mathrm{PCE}(t)$ trajectories reconstructed from LumPerNet predictions on the held-out test set (point-wise inference at acquisition time, not forecasting).
    Panels (a)--(c) show three devices sharing the same Architecture (A), selected to represent distinct degradation dynamics within our aging dataset:
    (a) relatively high stability with slow decay,
    (b) intermediate stability with moderate decay, and
    (c) relatively low stability with fast decay.
    Measured $R_\mathrm{PCE}(t)$ from indoor J--V characterization (pink) is compared to the LumPerNet ensemble-mean prediction (teal), obtained by averaging predictions from the four cross-validation trained models.
    At each aging time $t$, $\hat{R}_\mathrm{PCE}(t)$ is obtained by evaluating LumPerNet on the corresponding multimodal images and the $t=0$ reference state.
    Shaded regions denote $\pm1$ standard deviation across models.
    The horizontal axis reports elapsed time from the start of the aging experiment for each device; luminescence acquisitions are not necessarily performed at uniform intervals, and trajectories may therefore exhibit irregular temporal spacing.}
    \label{fig:trajectories}
\end{figure}

To complement aggregate metrics and parity plots, we qualitatively assess whether point-wise LumPerNet predictions yield temporally reconstructions of degradation trajectories of individual devices.
Figure~\ref{fig:trajectories} reports three representative trajectories from the held-out test set, covering distinct aging behaviors observed in the dataset.
Importantly, LumPerNet does \emph{not} forecast $R_\mathrm{PCE}$ from earlier data.
Instead, each point in the trajectory drawn using LumPerNet is obtained independently from the luminescence images acquired at the corresponding aging time $t$ (together with the device-specific reference images at $t=0$, i.e., by evaluating the trained regressor on $\mathbf{x}(t)$.
Thus, the trajectory is \emph{reconstructed} by repeating the same inference step for all available acquisition times of the device and plotting $\hat{R}_\mathrm{PCE}$ versus elapsed time.
For each device, the measured $R_\mathrm{PCE}$(t), derived from indoor J--V characterization, is shown alongside the ensemble-mean prediction obtained by averaging the outputs of the four CV--trained LumPerNet models; the shaded band indicates $\pm 1$ standard deviation across models at each acquisition time.

Across these examples, reconstructed trajectories follow the overall evolution of efficiency retention and remain within the operational range of interest.
Short-lived excursions in the measured $R_\mathrm{PCE}$ (e.g., the transient spikes around \mbox{$t\approx\SI{30}{\hour}$} in the left and central panels) are not reproduced by the model, consistent with the fact that they may reflect transient electrical effects or measurement artifacts of inconsistent power supply to the incident LEDs rather than persistent changes in luminescence patterns.
In addition, some devices exhibit systematic deviations between prediction and measurement, such as an overestimation during extended low-$R_\mathrm{PCE}$ plateaus (right panel), highlighting challenging regimes that may require larger datasets and, potentially, explicit handling of rapid early transients.
Overall, these trajectory-level examples support the quantitative benchmarks by showing that LumPerNet captures the dominant degradation trends while providing a natural measure of inter-model uncertainty through ensemble dispersion.

\begin{figure}[t]
    \centering
    \includegraphics[width=0.55\linewidth]{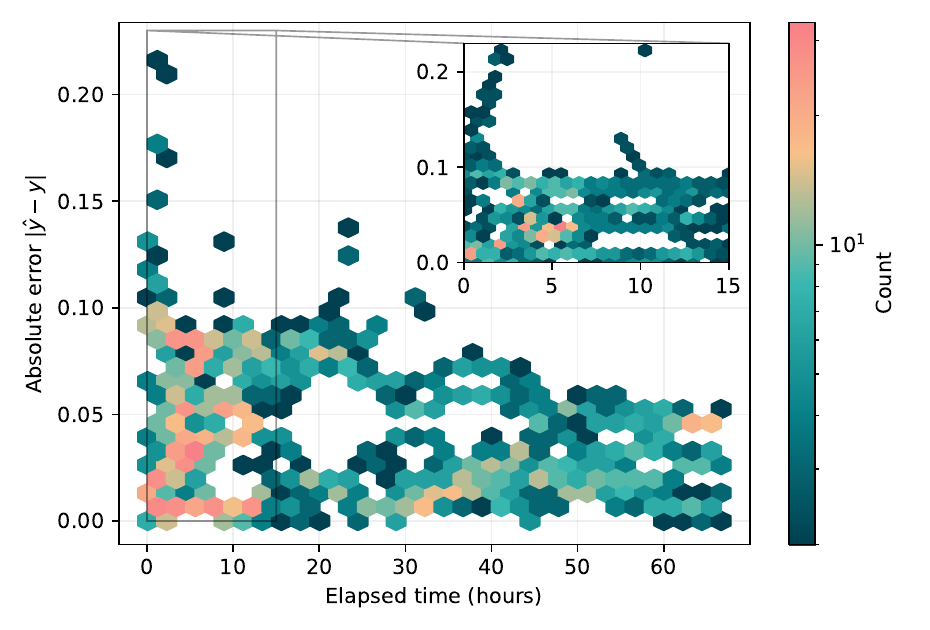}
    \caption{
    Absolute prediction error versus elapsed time on the held-out test set.
    Each bin reports the density of test-set samples in the $(t, \lvert \hat{R}_{\mathrm{PCE}} - R_{\mathrm{PCE}} \rvert)$ plane, where $t$ is the elapsed time from the start of aging for each device.
    The inset highlights the early-time regime (0--\SI{15}{\hour}), where measurements are most densely sampled and transient effects can lead to increased dispersion in the absolute error.
    }
    \label{fig:abs_error_vs_time}
\end{figure}

To complement the trajectory-level visualization in Figure~\ref{fig:trajectories}, we also quantify how prediction errors are distributed over time across the entire held-out test set.
Figure~\ref{fig:abs_error_vs_time} reports the absolute error $\lvert \hat{R}_{\mathrm{PCE}} - R_{\mathrm{PCE}} \rvert$ as a function of elapsed aging time, pooling all test-set samples.
This representation makes it possible to identify time windows associated with broader error distributions, and to distinguish isolated high-error events from systematic trends, without relying on a small number of selected devices.
Overall, the error distribution remains concentrated at low absolute errors across the majority of the experiment duration, while a broader spread is observed at early times, where rapid transients and reversible effects can induce short-term departures between predicted and measured $R_{\mathrm{PCE}}$.
The inset provides a detailed view of the first \SI{15}{\hour}, emphasizing that the early-time regime is both the most densely sampled and the most heterogeneous in terms of absolute error.

\subsection{Limitations and robustness considerations}\label{ssec:limitations}

While the results above demonstrate the feasibility of regressing PSC state-of-health from multimodal luminescence imaging, the present study remains a proof of concept and has several limitations that contextualize the reported performance.
First, the dataset spans 76 devices acquired across four experimental batches.
Three batches correspond to the same device architecture measured in different campaigns, whereas the remaining batch uses a different device architecture (composition and contact stack).
This heterogeneity, combined with the limited number of devices relative to the number of sampled time-points, contributes to fold-to-fold variability and motivates cautious interpretation of aggregate metrics.

Second, model predictions are constrained to an operational $R_\mathrm{PCE}$ window ($R_\mathrm{PCE} \in [0.8, 1.2]$) and are trained on samples within this range.
As a result, the current framework is not designed to reliably extrapolate to severe end-of-life degradation or to regimes dominated by early transient efficiency gains.
Extending the approach to full-lifecycle monitoring will require both broader datasets and explicit handling of out-of-window samples.

Third, the framework was intentionally kept lightweight and was not subjected to extensive hyperparameter optimization or architecture search.
Although LumPerNet slightly improves the fold-averaged held-out metrics and reduces fold-to-fold variability relative to the spatially homogenized control under identical splits, the performance gap is modest.
Further gains may be possible by incorporating richer temporal context (beyond reference--current pairing), improved photometric calibration (e.g., flat-field correction), and additional modalities or physics-informed constraints.

Finally, the batch-conditioning ablation highlights the presence of batch-structured variability in the dataset.
Explicit batch conditioning does not robustly improve LumPerNet and increases fold-to-fold variability, indicating that the spatial luminescence representation already absorbs part of the batch-dependent information available in the images.
This supports the use of the batch-free LumPerNet as the main model.
Importantly, however, all reported test devices are drawn from the same global batch pool used for model development.
The present results therefore demonstrate generalization to unseen devices under a device-level split, but not transfer to entirely unseen batches or architectures.
Such transfer is a more demanding domain-shift problem, particularly for PSCs, where nominally similar fabrication batches can still differ due to limited process repeatability~\cite{baumann_repeatable_2024,goetz_challenge_2022,zhang_achieving_2020}.
Dedicated leave-batch-out and prospective validation experiments are therefore required before deployment across new production domains.

Overall, these considerations suggest that the primary value of the present work is methodological: it establishes a reproducible, leakage-aware pipeline for coupling time series of luminescence images with electrical labels and demonstrates that spatially resolved multimodal information provides a modest robustness gain beyon a strong spatially homogenized control.
A natural extension is a two-stage formulation in which a lightweight classifier first flags samples outside the operational $R_\mathrm{PCE}$ window, followed by a regressor specialized for the range of interest.

%
%
\section{Conclusions}\label{sec:conclusions}

This work introduced a deep-learning framework for estimating the $R_\mathrm{PCE}$ of PSCs from multimodal luminescence imaging acquired during aging.
By pairing EL, PL\textsubscript{oc}, and PL\textsubscript{sc} images of an aged device state with device-specific reference images, the proposed CNN-based model (LumPerNet) learns spatial degradation signatures and directly regresses the normalized efficiency retention, defined according to Equation~\ref{eq:r_pce}, within an operational window relevant to stability testing.

Under a leakage-aware evaluation protocol based on device-level held-out testing and cross-validated training, LumPerNet achieves proof-of-concept predictive accuracy in the range $R_\mathrm{PCE} \in [0.8,1.2]$, and qualitatively tracks time-evolving degradation trends across representative devices.
Benchmarking against a spatially homogenized LumPerNet control shows that global luminescence evolution already contains substantial predictive information.
The homogenized control achieves $\mathrm{MAE} = 0.049 \pm 0.009$, $\mathrm{RMSE} = 0.065 \pm 0.015$, and $R^2 = 0.473 \pm 0.253$, compared with $\mathrm{MAE} = 0.048 \pm 0.005$, $\mathrm{RMSE} = 0.061 \pm 0.006$, and $R^2 = 0.558 \pm 0.078$ for the full spatial LumPerNet.
Thus, the benefit of spatially resolved inputs is moderate in terms of average error, but is reflected in improved fold-to-fold stability.
Modality-ablation experiments further clarify the role of multimodality: the full EL+PL\textsubscript{oc}+PL\textsubscript{sc} input yields the best generalization, while the single-modality models show a consistent drop in accuracy and cross-fold stability.
Importantly, the bimodal results are not monotonic with channel count: EL+PL\textsubscript{sc} retains most of the full-model performance, and PL\textsubscript{oc}+PL\textsubscript{sc} is competitive despite the weaker performance of PL\textsubscript{sc} in isolation, whereas EL+PL\textsubscript{oc} is markedly less stable across device splits.
Together, these findings support multimodal imaging as a means to capture complementary degradation signatures, and indicate that robustness depends on physically complementary contrast rather than simply adding additional channels.
Batch-conditioning ablations further show that part of the observed $R_\mathrm{PCE}$ variability is batch-structured: explicit batch metadata does not robustly improve LumPerNet.
This supports the choice of the batch-free LumPerNet as the default model, since it provides the best-performing and more transferable formulation without relying on dataset-specific identifiers.
Nevertheless, generalization to entirely unseen fabrication or measurement batches remains an important next step for deployment-oriented validation.

Beyond the specific results reported here, the primary contribution of this work is methodological.
The study establishes an end-to-end pipeline linking automated luminescence imaging to indoor electrical labels, with consistent preprocessing, training-time augmentation, and target-distribution balancing, and provides a baseline for future work on scalable, non-invasive degradation monitoring.
On the experimental side, future iterations of the acquisition platform will benefit from tighter environmental control (in particular relative-humidity monitoring and control) and improved photometric calibration, to further reduce measurement-induced variability and strengthen dataset consistency.
On the modeling side, future efforts will focus on expanding the dataset across device designs and aging conditions and extending the framework toward full-lifecycle monitoring.
A promising direction is a two-stage approach that first detects samples outside the operational $R_\mathrm{PCE}$ range and then applies a dedicated regressor within the region of interest, enabling reliable deployment in accelerated testing and, ultimately, quality-control settings.
More broadly, several complementary directions could be explored toward the same goal of scalable degradation monitoring, including hybrid formulations that combine the purely data-driven model presented here with device-physics descriptors, chemical degradation models, and ion-drift dynamics.
These include hybrid formulations that combine learned spatial features with compact physical descriptors or equivalent-circuit parameters, physics-informed or constraint-based learning to enforce monotonicity and operational bounds, and the integration of additional non-invasive channels (for instance temperature or impedance proxies) when available.
These outlooks align with the growing adoption of deep learning for image-based monitoring tasks in energy applications~\cite{casini_current_2024}.

\section*{Acknowledgements}\label{sec:acknowledgements}
S.T. acknowledges the European Union's Framework Programme for Research and Innovation Horizon Europe (2021–2027) under the Marie Skľodowska-Curie Grant Agreement No. 101107885 ``INT-PVK-PRINT''.
A.D.C. and S.T. acknowledge the European Union's Framework Programme for Research and Innovation Horizon Europe (2021–2027) under the Grant Agreement No. 101147311 ``LAPERITIVO''.
E.C. acknowledges partial support under the National Recovery and Resilience Plan (NRRP), Mission 4 Component 2 Investment 1.3-Call for tender No. 1561 of 11.10.2022 of Ministero dell'Università e della Ricerca (MUR), funded by the European Union NextGenerationEU.
We acknowledge AIFAC (project AIFAC\_S02\_113) for granting access to the LEONARDO supercomputer, owned by the EuroHPC Joint Undertaking, hosted by CINECA (Italy).

\section*{CRediT authorship contribution statement}\label{sec:author-contributions-statement}
\textbf{GB}: Conceptualization, Data curation, Formal Analysis, Investigation, Methodology, Software, Visualization, Writing -- original draft;
\textbf{ST}: Conceptualization, Data curation, Investigation, Methodology, Writing -- review \& editing, Supervision;
\textbf{SA}: Resources, Writing -- original draft;
\textbf{ZA}: Resources, Writing -- original draft;
\textbf{CO}: Resources, Writing -- original draft;
\textbf{MDA}: Resources, Writing -- original draft;
\textbf{EM}: Resources, Writing -- original draft;
\textbf{PA}: Conceptualization, Funding acquisition, Project administration, Supervision, Writing -- review \& editing;
\textbf{EC}: Conceptualization, Funding acquisition, Project administration, Resources, Supervision, Writing -- review \& editing;
\textbf{ADC}: Conceptualization, Funding acquisition, Project administration, Resources, Supervision, Writing -- review \& editing.

All authors read and approved the final manuscript.

\section*{Additional information}\label{sec:additional-information}
Supplementary Information is available with this article and includes measured LED emission spectra, photographs of the custom-built acquisition setup, detailed model architectures and hyperparameters, and extended results (complete tables, per-fold parity plots and ablation studies).

Portions of the wording in this work were refined with the assistance of ChatGPT, an AI language model by OpenAI, in accordance with the CC-BY 4.0 license; the underlying content was developed by the authors.

\section*{Conflict of interest}
The authors declare no conflicts of interest.

\section*{Code availability}\label{sec:code-availability}
The entire workflow is managed through Python scripts provided and maintained in the GitHub repository (\url{https://github.com/giuliobarl/LumPerNet}).
The repository includes the scripts used for image preprocessing, dataset assembly, model definition, training, and evaluation, as well as qualitative investigation of degradation trajectory predictions.

\section*{Data availability}\label{sec:data-availability}
The processed dataset used to train the models presented in this article are publicly available in the Zenodo repository (\url{https://doi.org/10.5281/zenodo.20157911}).
The Zenodo archive includes the processed luminescence image data, device/time metadata, $R_\mathrm{PCE}$ labels derived from synchronized J–V measurements, the exact per-fold file used to define the device-level train/validation/test partitions, the trained per-fold model checkpoints, and the generated figures.
These materials allow the reported leakage-aware evaluation protocol, fold-resolved predictions, and main quantitative results to be reproduced.

\section*{Competing interests}\label{sec:interests}
The authors declare no competing financial interest.


\newcommand{\supplementfilename}{supplementary.pdf}
\newcommand{\numbersupplementpages}{12} 

\includepdf[pages=-]{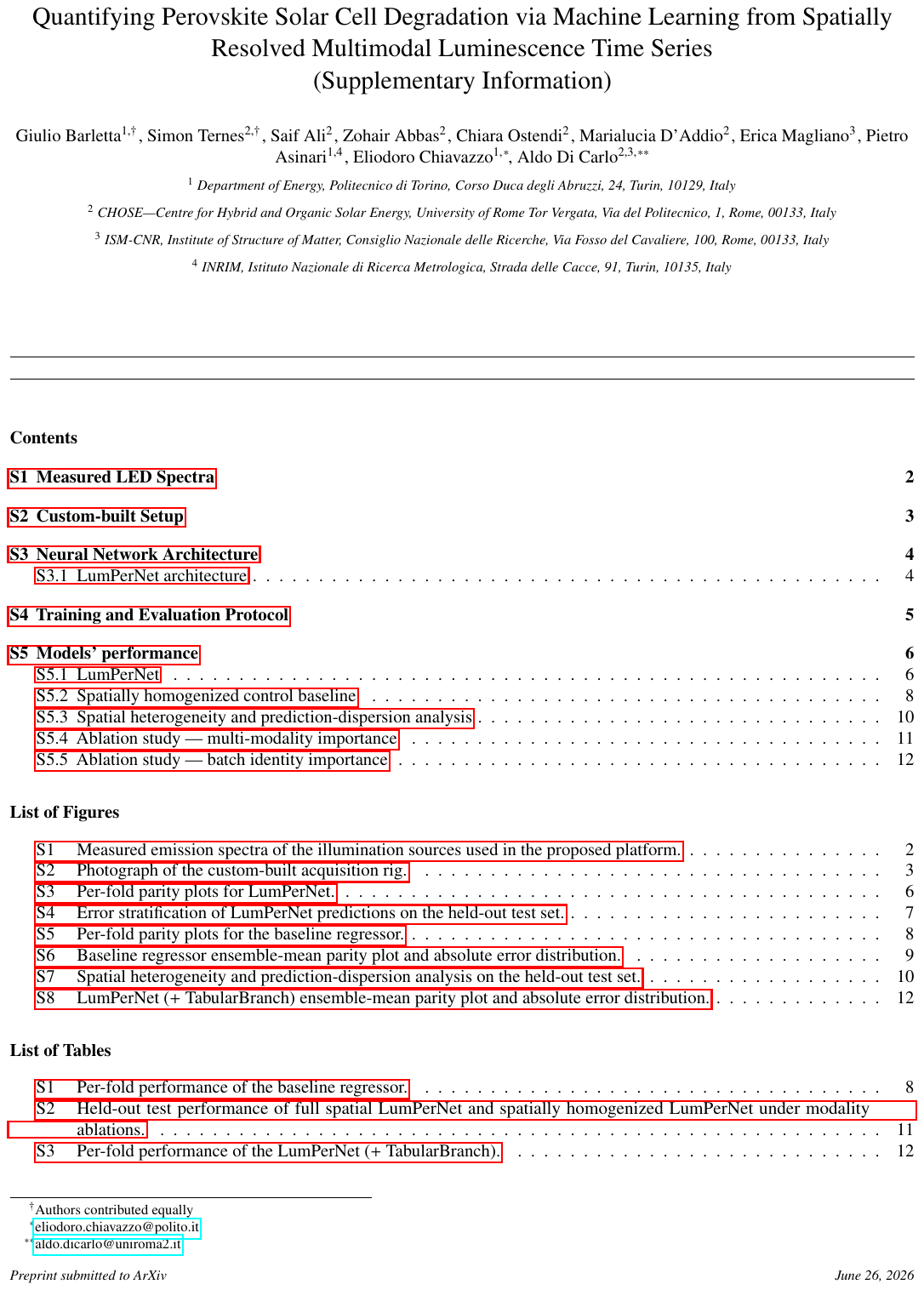}

\end{document}